\documentclass[
    reprint,
    superscriptaddress,
    amsmath,amssymb,
    aps,
    prx,
    floatfix,
]{revtex4-2}

\usepackage{graphicx} 
\usepackage{xcolor}
\usepackage{siunitx}
\DeclareSIUnit{\torr}{Torr}
\DeclareSIUnit{\rpm}{rpm}
\usepackage{braket}
\usepackage{bm}
\usepackage[colorlinks=true,linkcolor=blue,citecolor=blue]{hyperref}
\usepackage[capitalise]{cleveref}

\begin{document}

\title{Structural control of two-level defect density revealed by high-throughput correlative measurements of Josephson junctions}

\author{Oliver F.~Wolff}
\email{owolff2@illinois.edu}
\author{Harshvardhan Mantry}
\affiliation{Department of Physics, The Grainger College of Engineering, University of Illinois at Urbana-Champaign, Urbana, IL 61801, USA}

\author{Rahim Raja}
\affiliation{Department of Material Science and Engineering, The Grainger College of Engineering, University of Illinois at Urbana-Champaign, Urbana, IL 61801, USA}

\author{Wei-Hsiang Peng}
\author{Kaushik Singirikonda}
\affiliation{Department of Physics, The Grainger College of Engineering, University of Illinois at Urbana-Champaign, Urbana, IL 61801, USA}

\author{Seungkyun Lee}
\affiliation{Department of Material Science and Engineering, The Grainger College of Engineering, University of Illinois at Urbana-Champaign, Urbana, IL 61801, USA}

\author{Shishir Sudhaman}
\affiliation{Department of Physics, The Grainger College of Engineering, University of Illinois at Urbana-Champaign, Urbana, IL 61801, USA}

\author{Rafael Gonçalves}
\affiliation{Materials Research Laboratory, The Grainger College of Engineering, University of Illinois at Urbana-Champaign, Urbana, IL 61801, USA}

\author{Pinshane Y.~Huang}
\affiliation{Department of Material Science and Engineering, The Grainger College of Engineering, University of Illinois at Urbana-Champaign, Urbana, IL 61801, USA}
\affiliation{Materials Research Laboratory, The Grainger College of Engineering, University of Illinois at Urbana-Champaign, Urbana, IL 61801, USA}

\author{Angela Kou}
\affiliation{Department of Physics, The Grainger College of Engineering, University of Illinois at Urbana-Champaign, Urbana, IL 61801, USA}
\affiliation{Materials Research Laboratory, The Grainger College of Engineering, University of Illinois at Urbana-Champaign, Urbana, IL 61801, USA}
\affiliation{Holonyak Micro and Nanotechnology Lab, The Grainger College of Engineering, University of Illinois at Urbana-Champaign, Urbana, IL 61801, USA}

\author{Wolfgang Pfaff}
\email{wpfaff@illinois.edu}
\affiliation{Department of Physics, The Grainger College of Engineering, University of Illinois at Urbana-Champaign, Urbana, IL 61801, USA}
\affiliation{Materials Research Laboratory, The Grainger College of Engineering, University of Illinois at Urbana-Champaign, Urbana, IL 61801, USA}
\affiliation{Holonyak Micro and Nanotechnology Lab, The Grainger College of Engineering, University of Illinois at Urbana-Champaign, Urbana, IL 61801, USA}
\affiliation{National Center for Supercomputing Applications, University of Illinois at Urbana-Champaign, Urbana, IL 61801, USA}

\begin{abstract}
    Materials defects in Josephson junctions (JJs), often referred to as two-level systems (TLS), couple to superconducting qubits and are a critical bottleneck for scalable quantum processors.
    Despite their importance, understanding the microscopic sources of TLS and how to mitigate them has remained a major challenge. 
    Here, we demonstrate a high-throughput, correlated approach to trace the microstructural origins of strongly-coupled TLS in Josephson circuits. 
    We assembled a massive dataset of TLS across 6,000 Al/AlOx/Al JJs and more than 600 atomic resolution transmission electron microscopy images. 
    We statistically link fabrication, microstructure, and TLS occurrence, revealing a strong correlation between Al electrode thickness, Al grain size, and TLS density.
    Correspondingly, we find a two-thirds reduction in TLS prompted by a change in electrode fabrication parameters. 
    These results demonstrate a robust, data-driven methodology to understand and control defects in quantum circuits and pave the way for significantly reducing TLS density.
\end{abstract}

\maketitle

\section{Introduction}\label{sec: intro}

Unpredictable materials defects that couple to circuit modes are ubiquitous in superconducting quantum devices and present a major limitation on qubit coherence~\cite{mcrae_materials_2020,siddiqi_engineering_2021,deleon_materials_2021,murray_material_2021,bland_millisecond_2025}.  
These defects are typically referred to as two-level systems (TLS) and their coupling to quantum circuits results in excess loss and decoherence.
Of critical importance are TLS that couple strongly to circuit modes, allowing the coherent exchange of energy between quantum circuit and a single defect~\cite{cooper_observation_2004,simmonds_decoherence_2004,martinis_decoherence_2005,ku_decoherence_2005,ashhab_decoherence_2006,kim_anomalous_2008b}; if such a coupling occurs, the affected qubit is typically rendered effectively unusable.
This impact is in contrast to weakly coupled defects that merely manifest as excess dielectric loss~\cite{macha_losses_2010,woods_determining_2019,mcrae_materials_2020,molina-ruiz_origin_2021}.
The poorly-understood strongly-coupled TLS, however, significantly reduce reliability, casting doubt on the scalability of superconducting quantum processors.

Strongly-coupled TLS --- simply referred to as TLS in what follows --- are thought to occur in amorphous materials and interfaces in and around the commonly used Al/AlO\textsubscript{x}/Al Josephson junction (JJ), such as the oxide tunnel barrier.
The strong interaction  that limits qubit performance arises from their strong electric-dipole coupling to the qubit.
Motivated by the goal of improving qubit quality and scalability, there have been intense efforts to determine the nature and origin of such defects~\cite{faoro_microscopic_2007, sendelbach_magnetism_2008, desousa_microscopic_2009,dubois_delocalized_2013,lisenfeld_observation_2015,muller_interacting_2015,lisenfeld_decoherence_2016,klimov_fluctuations_2018a, lisenfeld_electric_2019, muller_understanding_2019, schlor_correlating_2019, bilmes_resolving_2020, bilmes_probing_2022,abdurakhimov_identification_2022, degraaf_chemical_2022, hung_probing_2022, carroll_dynamics_2022a, wang_spectroscopy_2025, colaozanuz_mitigating_2025, weeden_statistics_2025}.
To date, however, the microscopic origins of TLS remain unexplained and it is unclear to what extent they can be controlled or eliminated through device fabrication methods.

Because the origins of TLS are unknown and we cannot predict TLS behavior, we have only a limited toolset for mitigating their detrimental effects. 
For that reason, a range of post-fabrication mitigation techniques have been employed, such as (randomly) shifting TLS frequencies by thermal cycling~\cite{colaozanuz_mitigating_2025}, or by \emph{in-situ} application of strain~\cite{lisenfeld_decoherence_2016} and electric fields~\cite{dane_performance_2025}.
These strategies, however, come at the cost of significantly increased device and control complexity and, in the case of thermal cycles, significantly increased experimental time.
Further, none of these strategies are readily scalable to larger multi-qubit devices. 
It is therefore highly likely that achieving scalable quantum devices will demand the ability to predict the behavior of TLS by understanding their origins or, at a minimum, decrease their density through improved fabrication methods.

We can shed light on the origin of detrimental, strongly-coupled TLS by developing a procedure for correlating defect densities with fabrication parameters and microstructural characteristics.
While junction microstructure~\cite{zeng_atomic_2015,fritz_correlating_2018,oh_correlating_2025} and TLS statistics in qubits~\cite{colaozanuz_mitigating_2025,weeden_statistics_2025} have been investigated independently, no link has yet been established between structural features and the density of strongly-coupled TLS;
correspondingly, no structural controls that reduce TLS occurrence have been uncovered.
The identification of correlations has remained challenging because it requires large, correlated datasets that  meet the following criteria.
For one, a significant number of TLS must be detected in cryogenic measurements of Josephson circuits to infer their likelihood of occurrence.
Second, statistics on a range of microstructural parameters need to be acquired using materials characterization methods, such as electron microscopy.
By correlating sufficiently large datasets of microstructure and TLS, it should be possible to resolve correlations, provide critical clues for identifying TLS, and develop approaches to control or eliminate their formation during device fabrication.

Here, we introduce and demonstrate a coordinated characterization and analysis workflow for establishing correlations between fabrication parameters, densities of TLS strongly coupled to quantum circuits, and junction microstructure in Al/AlO\textsubscript{x}/Al JJs.
Using cryogenic measurements of JJ array resonators, we efficiently detect hundreds of distinct, strongly-coupled TLS, allowing us to extract their occurrence distributions for different fabrication recipes.
We examine the microstructure of the same junction arrays using scanning transmission electron microscopy (STEM), from which we extract features such as grain size, interface roughness, and layer morphology. 
We link TLS density to these microstructural characteristics and uncover a clear correlation between Al electrode thickness, Al grain size, and TLS densities.
Correspondingly, by modifying the Al electrode deposition methods we are able to prompt a significant reduction in TLS.
This identified correlation shows that reducing TLS density is possible by altering device fabrication methods, paving the way for significantly more reliable and scalable superconducting qubit devices.

\section{Strategy for identifying correlations}\label{sec: loop approach}

\begin{figure}
    \centering
    \includegraphics{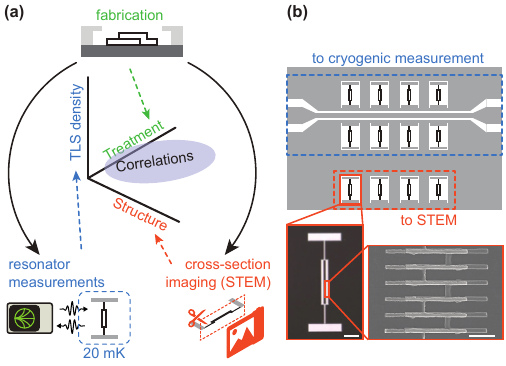}
    \caption{Resonator fabrication, measurement, and structural analysis loop.
    (a) For each fabrication treatment, we perform microwave measurements and microstructural characterization to extract TLS densities and grain-structure features at the junction interfaces.
    Combining these measurements with statistical inference allows us to identify correlations between fabrication treatment parameters, junction microstructure, and TLS density.
    (b) To ensure that the resonators used for TLS-density measurements and those used for microstructure analysis experience identical fabrication histories, both sets have the same design and are patterned on the same chip.
    Left scale bar: \qty{60}{\micro\meter}.
    Right scale bar: \qty{2}{\micro\meter}.
    Devices designated for TEM analysis are diced from the chip, while the remaining resonators are packaged for cryogenic rf measurements.
    }
    \label{fig: loop}
\end{figure}

To establish a workflow that enables the identification of statistically significant correlations between fabrication parameters, TLS densities in JJs, and JJ microstructure, we developed  a high-throughput, correlative fabrication and measurement workflow that can generate large datasets of TLS and microstructural data.
This approach is centered around the development of a series of methods that are specifically targeted for producing the required datasets [\cref{fig: loop}(a)]:
First, we employ junction-array resonators designed for maximizing the number of TLS coupling to them.
Second, we have realized a highly efficient and fully automated measurement and analysis workflow for inferring TLS densities from cryogenic circuit characterization.
Finally, we have devised a materials characterization and statistical correlation strategy that links microstructural features to fabrication variations and observed TLS densities.

We maximize the number of detectable defects by choosing frequency-tunable resonators containing JJ arrays, an approach chosen specifically for the purpose of uncovering correlations with statistical significance:
By utilizing arrays of large-area JJs, we can expect a large number of TLS strongly coupling to the resonator mode; flux-tunability ensures that we can tune the frequency of the resonators to resonance with a significant number of them.
In addition, we expect to be sensitive mainly to TLS associated with the oxide tunnel barrier of the JJs.
While there a range of areas in and around JJs that host TLS~\cite{lisenfeld_electric_2019,weeden_statistics_2025, zeng_atomic_2016}, defects in the junction are particularly important. 
First, the oxide barrier is a fundamental component of Josephson devices and TLS in the barrier can thus not be circumvented easily, unlike TLS in the leads, for example~\cite{weeden_statistics_2025}.
Second, their origin is unclear and the only method known to date to minimize the number of TLS in the barrier region is to shrink JJ area~\cite{colaozanuz_mitigating_2025}.

Based on prior estimates of TLS densities, $\rho_\text{TLS} \approx \qty{1}{\per\giga\hertz\per\micro\meter\squared}$ in JJs~\cite{bilmes_probing_2022, stoutimore_josephson_2012, martinis_decoherence_2005, palomaki_multilevel_2010a, colaozanuz_mitigating_2025}, each array is designed to include approximately \qty{100}{\micro\meter\squared} of total junction area, ensuring a substantial number of detectable TLS per resonator; in contrast, in methods employing transmon qubits, with typical junction area $< \qty{1}{\micro\meter\squared}$, orders-of-magnitude more devices need to be probed to yield the same statistics.
Each chip incorporates eight such resonators, frequency-multiplexed along a shared feedline, allowing us to accumulate statistics with minimal complexity and in a single cooldown [\cref{fig: loop}(b)]. 
Each such device corresponds to a unique set of fabrication parameters; 
varying fabrication processes between devices then enables systematic study of their impact on TLS densities.
Device fabrication details are given in \cref{supp: fab}.

To efficiently capitalize on the large JJ area per device, we developed a cryogenic measurement and analysis workflow that is designed to detect TLS at cryogenic temperatures and derive bounds on the TLS density.
Each resonator is measured using a custom-developed fully-automated data acquisition system.
The system enables fast sequential measurements of the resonator response as a function of its external flux-tuned frequency without manual intervention.
From these resonator measurements we automatically extract signatures of strong TLS coupling directly from resonator spectroscopy data, a significant simplification over methods relying on qubit control.
The gain in efficiency can be illustrated by comparison:
As shown below, we have detected 393 distinct TLS across 5 samples, detected with simple resonator spectroscopy. 
For comparison, in one of the largest TLS datasets published so far, \citet{colaozanuz_mitigating_2025} used qubit swap spectroscopy, a significantly more complex pulsed measurement protocol, to detect 109 TLS across 20 samples containing a total of 92 qubits.

Having detected a large number of TLS is this way, we then employed an empirical-Bayesian model to infer uncertainty bounds on the TLS density for each device, and thus each distinct fabrication variation.
Details on the cryogenic measurement and analysis workflow are presented in \cref{sec: density extraction}. 
Extracted TLS densities and their variations across different devices are discussed in \cref{sec:tls-densities}.

Our materials characterization focuses on analyzing the microstructure of each device variation with scanning transmission electron microscopy (STEM). 
As illustrated in \cref{fig: loop}(b), the same fabrication run yields devices with the same geometry for both cryogenic measurements and microstructural analysis to ensure consistency.
STEM images of the junction cross-sections allow us to extract key structural features, such as lateral grain size, junction roughness, and electrode roughness. 
The data produced by imaging is thus a catalog of microstructural properties, taken with sufficient statistics for each fabrication variation.
Details on the measured properties are presented in \cref{sec: TEM}.

Finally, the cryogenic measurements and microstructural analysis feed into a statistical analysis for identifying correlations between fabrication parameters, junction microstructure, and TLS density. 
This analysis aims to identify fabrication ``knobs'' that influence TLS density to provide crucial insights into the physical origins of these correlations. 
In \cref{sec: grain stats}, we present the methods for building these correlations and interpreting their significance, which are critical for optimizing fabrication processes to control TLS prevalence in quantum devices.

\section{Extraction of TLS density from cryogenic measurements}\label{sec: density extraction}

\begin{figure*}[t!]
    \centering
    \includegraphics{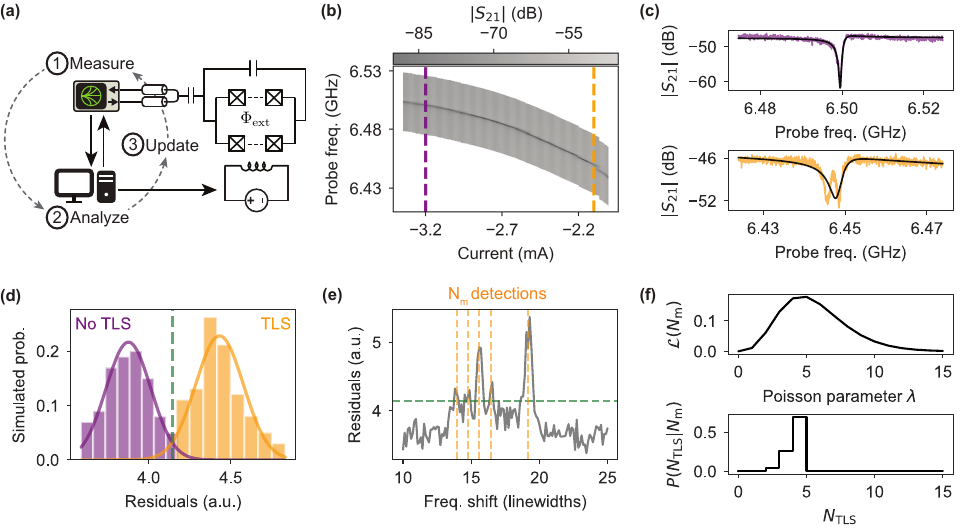}
    \caption{TLS detection workflow.
    (a) Resonator responses are measured and fit to the hanger model to extract resonator frequency, $f_0$, and residuals.
    The center frequency and applied d.c. flux bias is updated automatically over a pre-defined current sequence.
    (b) Example data produced by this autonomous `curve-following' measurement procedure.
    The yellow line cut exhibits a clear avoided crossing (AvC).
    (c) Resonator responses marked by dashed lines in (b), with absence (purple) and presence (yellow) of TLS coupling. 
    Black lines are fits to the standard `hanger'-model.
    (d) Simulated residual distributions used to calibrate the detector.
    Each resonator is simulated with (yellow) and without (purple) TLS coupling.
    Gaussian fits to the two distributions are used to determine the optimal detection threshold (in green), corresponding to the threshold in (e).
    (e) Residuals from a curve-following measurement, with tuning-axes re-normalized to units of resonator linewidth.
    The calibrated TLS-detection threshold is indicated, along with the identified TLS events.
    (f) \textit{Top}: Poisson likelihood function $\mathcal{L}(\lambda)$ for the TLS density parameter $\lambda$, conditioned on the number of detections $N_\text{m}$.
    \textit{Bottom}: Posterior probability distribution for the number of TLS in the measured sample.
    The inference incorporates the number of detections from (e), the false-positive and false-negative rates determined in (d), and the maximum-likelihood estimate of $\lambda$ inferred from $\mathcal{L}(N_\text{m})$.}
    \label{fig: detection}
\end{figure*}

We measured resonator responses at cryogenic temperatures to identify strongly-coupled TLS in our JJ array resonators.
To efficiently accumulate statistics we employed a simple protocol that achieves automatic TLS identification with minimal control complexity. 
In contrast to methods that utilize qubit control for the detection~\cite{weeden_statistics_2025,lisenfeld_decoherence_2016, colaozanuz_mitigating_2025, carroll_dynamics_2022a, klimov_fluctuations_2018a}, we developed a simple approach based on resonator spectroscopy. 
We acquired resonator response traces vs.\ frequency, $S_{21}(f)$, over a range of externally applied flux values, corresponding to the frequency-tuning range of the resonator.
These measurements are performed fully automatically, and implemented using an efficient `curve-following' program [\cref{fig: detection}(a,b)]. 
Details on the measurement setup are given in \Cref{supp: resonator measurement}.

To identify likely TLS, we inspect the residuals of a fit to the measured resonator response, $\sigma^2(\text{data} - \text{fit})/\mu(\text{data})$, where $\sigma^2$ is the variance of the difference between data and fit, and $\mu$ is the mean of the data.
In the presence of a strongly coupled TLS, the standard `hanger' resonator model~\cite{khalil_analysis_2012} fails to capture the response accurately [\cref{fig: detection}(c)], producing sharp, localized peaks in the fit residuals which serve as indicators of TLS couplings that can be identified without manual inspection.

While this detection method is extremely efficient, it is naturally prone to errors from non-ideal resonator responses.
Before beginning a measurement, we use a single-photon power measurement of the resonator to calibrate the optimal sampling rate, bandwidth, and integration time to mitigate non-ideal resonator responses and detrimental measurement noise.
Measurement noise and fitting errors mean that some true TLS fall below the detection threshold (false negatives), while spurious features could be flagged as TLS (false positives).
To quantify these effects and extract the best-estimate TLS occurrence, we calibrate our detection method using synthetically generated reference data.

The reference data are generated by performing simulations that replicate the behavior of each measured resonator, in both the absence of and presence of TLS coupling [\cref{fig: detection}(d)].
We simulate minimally detectable TLS with cooperativity $C=4g^2/\kappa\gamma=1$ for coupling strength $g = g_{\mathrm{crit}}=\kappa/2$, TLS decay rate $\gamma = \gamma_{\mathrm{crit}}=\kappa$, and detuning from the resonator $\omega_\text{TLS}-\omega_\text{resonator} = \kappa/2$ (see \Cref{supp: calibration} for more details).
The distributions of both baseline and TLS-coupled residuals are broadened by measurement noise, resulting in Gaussian-distributed histograms [\cref{fig: detection}(d)].
The intersection of these distributions defines a detection threshold: above this value, a residual is more likely to arise from a TLS than from noise.
The false positive rate for the detector is then computed as the integrated area under the noise-only Gaussian above this threshold, while the false negative rate is given by the area under the TLS-coupled Gaussian below the same threshold.
In practice, TLS will couple to resonators with different $g$ and have different $\gamma$. 
By calibrating to the minimally detectable TLS, where the overlap between noise-only and TLS-coupled residuals is maximal, the resulting false positive and false negative rates represent worst-case (upper-bound) errors.
TLS with stronger coupling ($g> g_\text{crit}$) and/or longer lifetimes ($\gamma<\gamma_\text{crit}$) are correspondingly easier to detect.
This procedure yields the false positive and false negative rates for a single trace under realistic noise conditions.

Finally, the outcomes of the detector calibration enable us to estimate probability distributions for the occurrence of detectable TLS (see \cref{supp: stats} for more information).
We assume that each TLS occurs independently and, when taken with our maximum 1 TLS$/\kappa$ detection resolution, construct a maximum likelihood Poissonian prior for the appearance of TLS in our devices [\cref{fig: detection}(f), top].
We then employ an empirical Bayes approach to construct the posterior distribution over the true number of detectable TLS given the number of detections [\cref{fig: detection}(f), bottom]. 
From this posterior, we extract both the expected TLS count and credible intervals, which provide statistical error bounds on the TLS density.
To compute the TLS density, $\rho_\text{TLS}$, we normalize the expected count, $N_\text{TLS}$, by the frequency range swept by the resonator, $\Delta f$, and the total junction area, $A$, such that $\rho_\text{TLS}=N_\text{TLS}/(\Delta f\cdot A)$.

\section{TLS density variations}\label{sec:tls-densities}

\begin{figure}
    \centering
    \includegraphics{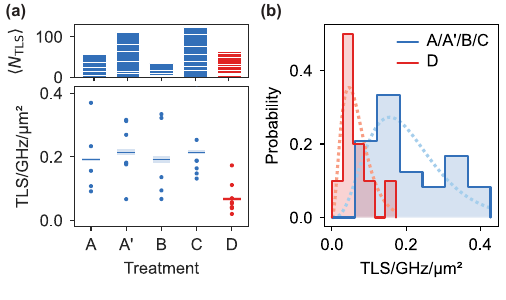}
    \caption{Effect of fabrication parameters and treatments on TLS densities.
    Key: A, A\textsf{'} – standard fabrication; B – annealed at \qty{250}{\degreeCelsius} for 20 minutes; C – slower aluminum deposition rate (\qty{0.6}{\angstrom\per\second}); D – thicker Al electrodes (bottom: \qty{100}{\nano\meter}, top: \qty{150}{\nano\meter}).
    (a) \textit{Top}: Extracted mean number of TLS, $\langle N_\text{TLS}\rangle$.
    Each bar is segmented by individual resonator contributions, grouped by fabrication treatment.
    \textit{Bottom}: TLS densities for different fabrication treatments.
    Dots represent individual resonator measurements. 
    Horizontal lines indicate the mean TLS density, and shaded rectangles denote the standard error of the mean.
    (b) Step plots of the TLS densities from (a), with overlaid scaled gamma fits (dotted curves) used to model the distributions.
    Thicker electrodes (red; $\rho_\text{TLS}=0.07\pm0.04$~TLS \unit{\per\giga\hertz\per\micro\meter\squared}) show a two-thirds reduction in TLS density compared to unannealed, annealed, and slow-deposition samples (blue; $\rho_\text{TLS}=0.20\pm0.10$~TLS \unit{\per\giga\hertz\per\micro\meter\squared}).}
    \label{fig: tls densities}
\end{figure}

Armed with a calibrated detector and statistical framework, we investigated correlations between fabrication treatment and $\rho_\text{TLS}$ by measuring multiple devices fabricated under varying conditions.
First, we measured a device that was made using a standard fabrication recipe (treatment A; see \cref{supp: fab} for details).
Systematic measurements with our detector reveal a broad spread in $\rho_\text{TLS}$ across nominally identical resonators [\cref{fig: tls densities}(a)], in line with previous results~\cite{bilmes_probing_2022}. 
This device establishes a baseline mean TLS density and a reference level of variability, serving as the control for evaluating the effects of subsequent fabrication modifications.
The variability of the number of detected TLS illustrates the importance of measuring multiple resonators to uncover the underlying probability distribution.
Individual resonator measurements provide only noisy samples from the parent distribution, and to accurately resolve the underlying distribution and draw reliable conclusions about the sources of TLS, a high throughput study is essential.

To test whether modest fabrication changes affect $\rho_\text{TLS}$, we fabricated additional devices with varied treatments: post-fabrication annealing at \qty{250}{\degreeCelsius} in an inert N$_2$ environment for 20 minutes (treatment B) and using a slower aluminum deposition rate of \qty{0.6}{\angstrom\per\second} (treatment C). 
These modifications were motivated by the following considerations:
Thermal annealing is known to reduce structural disorder, promote recrystallization, and improve interfacial oxide stoichiometry, all of which may suppress defect formation in the junctions~\cite{korshakov_aluminum_2024, koppinen_complete_2007}.
Changing the aluminum deposition rates can alter the grain size and film roughness~\cite{oh_correlating_2025}, potentially altering the density of TLS hosted at grain boundaries.
To test reproducibility we added another control, treatment A\textsf{'}; this device is identical to treatment A except for post-fabrication aging, an effect we expect to have minimal impact on $\rho_\text{TLS}$.
These treatments produced similar means and spreads, suggesting that none of the fabrication modifications resulted in a noticeable change in $\rho_\text{TLS}$. 

Finding little variation in mean $\rho_\text{TLS}$ across these first four devices, we then turned to a fabrication change that has previously been shown to suppress power-dependent, typically TLS-associated loss in CPW resonators: increasing the aluminum electrode thickness~\cite{biznarova_mitigation_2024}.
This prior work focused on loss sources near metal-substrate interfaces that reduce quality factors. 
In contrast, our aim is to test whether the aluminum thickness of junction electrodes affects the density of TLS strongly coupled to JJ devices.
We fabricated devices with thicker Al layers (treatment D: bottom electrode \qty{100}{\nano\meter}, top electrode \qty{150}{\nano\meter}) to test whether this change would reduce $\rho_\text{TLS}$.
Ensemble measurements with our detector confirmed this expectation: both the mean and variance of $\rho_\text{TLS}$ are noticeably lower than in devices from treatments A–C [\cref{fig: tls densities}(a)].

We then applied a series of statistical tests to directly compare the distributions associated with each treatment and rigorously establish this observed difference.
Devices A-C were not statistically distinguishable, while device D was statistically distinguishable from all others.
We modeled the combined A-C distributions with a gamma distribution to accommodate the skewness and positivity of the $\rho_\text{TLS}$ data. The resulting fit yielded a mean $\rho_\text{TLS}$ of $0.20 \pm 0.10$ \unit{\per\giga\hertz\per\micro\meter\squared}.
Device D, in contrast, exhibits a significantly lower TLS density (Kruskal-Wallis $p \leq 0.05$) with a gamma-fitted mean of $0.07 \pm 0.04$ \unit{\per\giga\hertz\per\micro\meter\squared} [\cref{fig: results}(b)].
This result corresponds to a two-thirds reduction compared to devices A–C.
We note that the observed TLS density is about a factor of 5 lower than what has been reported in other recent studies~\cite{colaozanuz_mitigating_2025,weeden_statistics_2025}.
Because of differences in device geometry and detection heuristic, absolute numbers can most likely not be compared directly.
The demonstration of a two-thirds reduction within a single self-consistent study with sufficient statistics, however, is strong evidence for the effect of a specific method in altering TLS density.
No such method has been previously been shown for reducing density of TLS associated with the tunnel barrier of the JJs.


\section{Structural analysis using TEM and correlation with TLS density}
\label{sec: TEM}
\label{sec: grain stats}

\begin{figure}
    \centering
    \includegraphics{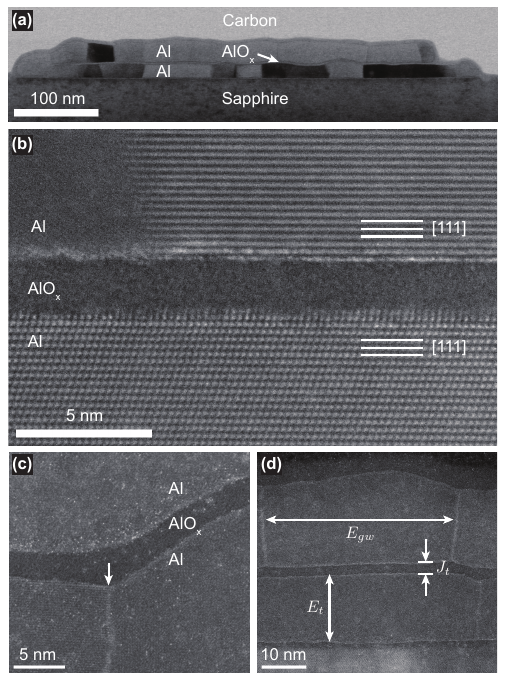}
    \caption{Cross-sectional STEM images of Josephson junctions.
     (a) BF-STEM image of one Josephson junction. The intensity variations correspond to different grains in the Al electrodes. 
     (b) ADF-STEM image of a flat, uniform junction with [111] facet facing the junction for both top and bottom electrodes.    
     (c) ADF-STEM image of enlarged junction width due to grain boundary grooving.
    (d) ADF-STEM image of Al/AlO\textsubscript{x}/Al showing junction morphology metrics measured. $E_t$ is the thickness of the bottom electrode. $J_t$ is the thickness of the junction as a function of position.  $E_{gw}$ is the lateral Al grain width.}
    \label{fig: stem}
\end{figure}

We performed detailed microscopic structural characterization across all samples to understand the possible origins of the reduced TLS density observed in thicker aluminum trilayers.
We prepared TEM samples from devices on the same wafer as those used for TLS measurements (\cref{supp: tem}). 
Resulting images are presented in \cref{fig: stem}. 
\cref{fig: stem}(a) shows a low magnification BF-STEM image of the junction structure, which consists of polycrystalline aluminum electrodes surrounding an amorphous AlO\textsubscript{x} junction. An atomic resolution image of a junction is shown in \cref{fig: stem}(b). These STEM images provide rich, multi-scale insight into the junction structure. For example, we  observe variations in the junction thickness, such as a local increase of AlO\textsubscript{x} thickness near Al grain boundaries (see \cref{fig: stem}(c)) due to grain boundary grooving~\cite{mullins_theory_1957,oh_correlating_2025}. 
The STEM images also reveal structural defects that could potentially be correlated with TLS. As shown in the atomic resolution image in \cref{fig: stem}(b), the Al/AlO\textsubscript{x} interfaces contain dense atomic steps and are rough on the angstrom scale. Variations in the Al/AlO\textsubscript{x} interfacial structure, Al grain boundary density, and junction morphology could each be correlated with TLS.

\begin{figure}
   \centering
   \includegraphics{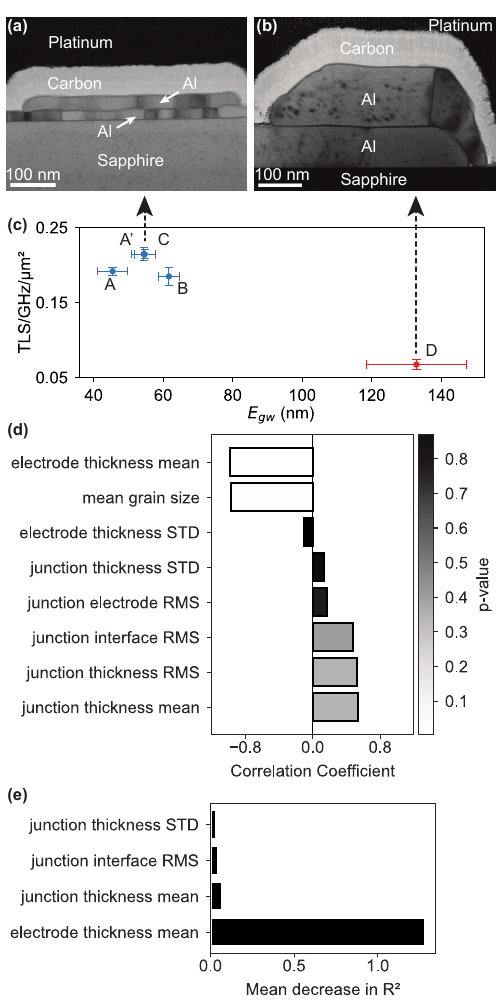}
   \caption{Correlation between TLS densities and various structural metrics extracted from STEM images.
    (a, b) Cross-sectional BF-STEM images of junctions from samples C and D, respectively, illustrating the grain structure.
   (c) Plot of TLS density versus lateral Al grain size. Sample D, which features thicker trilayers, exhibits both significantly reduced TLS densities and larger grains compared to samples A, A\textsf{'}, B, and C. Each data point represents a sample average. Vertical error bars correspond to the standard error in TLS density (as in \cref{fig: tls densities}), and horizontal error bars indicate the 68.27$\%$ confidence interval for the mean lateral grain size.
   (d) Pearson correlation coefficients and associated P-values of measured STEM features against mean TLS density. Only the electrode thickness mean and mean grain size are statistically significant. 
   (e) Ranking of feature impact on Ridge regression model performance from permutation importance analysis. The mean electrode thickness is shown to be the most important factor to predict TLS from STEM images. }
   \label{fig: results}
\end{figure}

Next, we developed a methodology to rapidly parameterize, quantify, and correlate key metrics of junction morphology. We collected a large dataset of 709 STEM images spanning 5 devices, with each device containing 6-10 junctions. In total, we characterized a lateral distance of 8.4 microns across all devices. This large set of images was needed to separate  systematic differences between different preparation methods from random variations in morphology, which were also present in our devices.  From our images, as illustrated in \cref{fig: stem}(d), we measured the Al electrode thickness ($E_t$), the lateral grain sizes of the electrodes ($E_\text{gw}$), and the AlO\textsubscript{x} thickness ($J_{t}$), and computed their mean, standard deviation (STD), and root mean square (RMS) roughness. 

As shown in \cref{fig: results}(a-c), we observe clear correlations between lateral grain size, electrode thickness, and TLS density.  Devices fabricated with treatments A, A\textsf{'}, B, and C, which we determined above to have higher TLS densities, each contain thin Al electrodes with mean grain sizes between 40-60 nm. Meanwhile, treatment D, which exhibits lower $\rho_\text{TLS}$, also has larger Al grains (mean of  128 nm).  We also observed differences in the orientations of these grains. For devices for treatments A-C, 90\% of the bottom electrode grains and 70 \% of the top electrodes exhibit [111] orientations towards the junction. All grains were faceted [111] towards the junction in the 26 grains imaged in device D. 
We used two statistical tests to assess the strength and importance of correlations. As shown in  \cref{fig: results}(d), we computed Pearson correlation coefficients  to measure the strength of pairwise correlation between STEM metrics and TLS density. Of the parameters measured, only two ---the mean electrode thickness and mean Al grain size--- show statistically significant ($p<0.05$) correlations with TLS density.  

In order to compare the relative impact of different morphological features, we use ridge regression to model the TLS density as a function of the metrics measured by STEM. 
We isolated four representative features: the junction thickness mean, junction thickness STD, junction interface RMS roughness, and the mean electrode thickness. We then computed the permutation importance using ridge regression for these four features. \Cref{fig: results}(e) compares the predictive influence of the representative metrics. 

From our ridge regression model, we found that the mean electrode thickness (which was clustered with lateral grain size mean), was the most important metric for predicting device TLS density, in agreement with our other statistical analyses. 
Although here, only one independent metric was identified to be correlated with TLS density,  similar analyses can be used to rank features in cases where multiple correlations are present. 

Taken together, our analyses reveal a strong correlation between thicker Al electrodes, larger Al grains, and reduced TLS density. These correlations imply that grain boundaries, or the AlO\textsubscript{x} adjacent to the grain boundaries, are sites for TLS formation. Moreover, our statistical analysis of large STEM datasets provides a systematic approach to a problem that has historically been challenging: identifying and correlating microstructural features with device properties in complex polycrystalline multilayers.

\section{Discussion}\label{sec: conclusion}
We have developed an efficient and statistically rigorous methodology for correlating the density of TLS detected in cryogenic measurements of Josephson circuits with fabrication methods and microstructure of these circuits.
We have fabricated devices that incorporate JJ arrays in multiple microwave resonators to identify a sufficiently large number of distinct, strongly-coupled TLS to resolve their probability distributions of occurrence. 
Importantly, we have confirmed that these distributions demand the measurement of ensembles of resonators to extract reliable, quantitative TLS densities.
In addition, we have systematically analyzed microstructural features of JJ arrays that were co-fabricated with the cryogenically measured devices.
The analysis of these datasets has yielded statistically significant correlations between fabrication procedure, structure, and TLS density.
Our methodology thus introduces a powerful approach for elucidating the origin of strongly-coupled TLS in Josephson quantum devices.

Using our workflow for establishing correlations, we systematically investigated  devices fabricated under different processing conditions and compared the resulting TLS densities.
Among the fabrication parameters tested, varying aluminum thickness produced the most striking effect: 
junctions with thicker electrodes exhibited significantly larger grains, fewer grain boundaries, and a two-thirds reduction in TLS density relative to standard fabrication treatments.
An exhaustive statistical analysis reveals mean grain size and mean electrode thickness to be the primary contributors to observed decrease in TLS density.
This strong correlation between grain morphology and TLS density highlights the microstructure of the junction as a dominant source of TLS and provides evidence that fabrication choices can substantially alter defect populations.

Our observations differ from prior studies which have found that TLS density in the oxide barrier is relatively insensitive to changes in JJ fabrication, leaving only junction area as a control knob~\cite{colaozanuz_mitigating_2025}. 
The uncovering of electrode thickness and grain size as control knobs is one of the central new insights of our work.
The observed correlations demonstrate that TLS density in Al/AlO\textsubscript{x} JJs are not fixed at an intrinsic limit, but can be substantially changed through minor modifications to device design and fabrication.
While more data will be needed to find the ultimate limits to which TLS can be suppressed across a range of different devices, a two-thirds reduction of TLS in Josephson qubits would already be a significant enhancement in average coherence and reliability of qubit devices.
Because the probability of a multi-qubit device to be fully functional is exponential in the number of qubits, such a reduction would already dramatically enhance the probability of a device with many qubits to not suffer from any malfunctioning qubits.
Taken together, our results point toward grain engineering in JJ fabrication as a promising strategy for reducing TLS occurrence and thus enhancing coherence and reliability in superconducting qubit devices.

\section*{Acknowledgments}
This research was carried out in part in the Materials Research Lab Central Facilities and the Holonyak Micro and Nanotechnology Lab, University of Illinois. 
This material is based upon work supported by the Air Force Office of Scientific Research under award numbers FA9550-23-1-0690 FA9550-25-1-0306.



\appendix

\section{Device fabrication}\label{supp: fab}
Junction-array resonator devices were fabricated on \SI{8}{\milli\meter}~$\times$~\SI{5}{\milli\meter} coupons diced from 2-inch C-plane sapphire wafers sputter-coated with \SI{200}{\nano\meter} of tantalum (purchased from {\em StarCryo}). 
Each coupon was subsequently split into two regions: a \SI{3}{\milli\meter}~$\times$~\SI{5}{\milli\meter} segment designated for TEM analysis and a \SI{5}{\milli\meter}~$\times$~\SI{5}{\milli\meter} chip used for cryogenic measurements, as shown in \cref{fig: loop}(b). 
In this way, both measurements probe junctions with the same geometry originating from the same fabrication run, thereby minimizing systematic microstructural differences.

The Ta ground plane, capacitor pads, and feedline were patterned using AZ-1518 photoresist. 
Exposure was performed using a Heidelberg MLA~150 maskless aligner (\SI{375}{\nano\meter} wavelength; \SI{80}{\micro\joule\per\centi\meter\squared} dose). 
The resist was developed in AZ 917 MIF and underwent a \SI{2}{\minute} oxygen descum at \SI{300}{\milli\torr} O$_2$ with \SI{18}{\watt} RF power. 
The patterned films were etched using Transene Tantalum Etchant 111 for \SI{17}{\second} while stirring at \SI{120}{\rpm}, followed by sequential \SI{30}{\second} DI water dunk and rinse.
Josephson junctions were fabricated using the bridge-free technique with double-angle evaporation~\cite{lecocq_junction_2011}. 
The resist stack consisted of MMA~EL13/PMMA~950k~A4 patterned by electron-beam lithography on a Raith EBPG5150 (\SI{100}{\kilo\volt}) and developed in a 3:1 IPA:DI mixture at \SI{5}{\celsius} for \SI{2}{\minute}. 
The aluminum trilayer was deposited via electron-beam evaporation in a Plassys MEB550S4II system: \SI{25}{\nano\meter} at +\SI{45}{\degree}, followed by \SI{80}{\minute} of static oxidation at \SI{60}{\torr}; then \SI{35}{\nano\meter} at $-\SI{45}{\degree}$, followed by \SI{5}{\minute} oxidation at \SI{10}{\torr} to form the capping layer.  
Both evaporation steps proceeded at \SI{3}{\angstrom\per\second}. 
Each resonator incorporated junctions of one uniform size, either \SI{0.32}{\micro\meter}~$\times$~\SI{5}{\micro\meter} or \SI{0.48}{\micro\meter}~$\times$~\SI{5}{\micro\meter}.

\section{Resonator measurement}\label{supp: resonator measurement}
Resonators were cooled to a base temperature of \SI{10}{\milli\kelvin} in an Oxford Triton 500 dilution refrigerator and enclosed within a copper sample box surrounded by aluminum and mu-metal magnetic shielding.
A standard configuration of microwave components as typically used for superconducting qubit-experiments, including cryogenic attenuators, circulators, and HEMT amplifiers, was used to connect the device to radio-frequency measurement equipment~\cite{krinner_engineering_2019b}.

JJ-array resonator responses were then acquired with a vector network analyzer (VNA).
Flux-tuning was achieved by driving a superconducting flux line with a Yokogawa GS200 current source.
The applied current generated a local magnetic flux threading the JJ array, thereby shifting its inductance and, by extension, resonance frequency.
At each flux-bias point, the VNA performed a continuous-wave frequency sweep of the complex transmission, $S_{21}(f)$, near single-photon power.

Each set of resonator data requires filtering to identify suitable regions for analysis and to determine which fits can be used to generate TLS-free synthetic data. 
Since each chip contains multiple resonators operating within the 4–8 GHz range, resonators may lie close in frequency or even overlap as the flux is swept. 
An example is shown in \cref{fig: filtering}(a), where a second resonator passes through the one being tracked. 
The hanger response model cannot account for multiple resonances, so the influence of a second nearby resonator, especially at close range, can be indistinguishable from that of a TLS. These regions are excluded from the analysis. 

Resonator frequencies exhibit approximately parabolic flux dependence, as shown in the equation below:
\begin{align}\label{eq: freq flux dependence}
    f(\Phi_\text{ext})&=\frac{f_R}{\sqrt{1+\frac12\left[\frac{2\pi}{N}\left(\frac{\Phi_\text{ext}}{\Phi_0}-m\right)\right]^2}}\\
    &\approx f_R\left(1-\frac{\pi^2}{N^2}\left(\frac{\Phi_\text{ext}}{\Phi_0}-m\right)^2\right),
\end{align}
where $\Phi_{ext}$ is the external applied flux, $f_R$ is the bare resonator frequency, $N$ is the number of superconducting islands in the array, $\Phi_0$ is the flux quantum, and $m$ is the number of flux quanta in the loop~\cite{masluk_microwave_2012}. 
The approximate quadratic relation holds when the external applied flux is close to the trapped flux quanta in the loop.
This dependence can result in TLS being detected twice when a resonator is tracked through its maximum frequency, leading to an artificial inflation of the TLS count. 
To prevent this over-counting, we exclude data beyond the point at which the resonator frequency has reached its maximum.

To reliably flag features in the fit residuals and robustly identify strongly-coupled TLS candidates we analyze the data as follows.
Because TLS appear as distinct peaks in the residuals, the resonator linewidth sets a fundamental limit on how closely they can be spaced in frequency and still be resolved.
This effective detector resolution permits the identification of at most one TLS per resonator linewidth, $\kappa$.
However, our data are acquired by sweeping the applied current, and since the resonator frequency depends quadratically on current when $\Phi_\text{ext}/\Phi_0\ll1$, fixed current steps do not correspond to uniform frequency steps.
As a result, identical resonator-TLS couplings can have varying residual peak widths depending on their position along the sweep.
We fit the resonator frequency to a quadratic function of the current, allowing us to compute the frequency shift of the resonator.
The spacing of residuals is then normalized to frequency shift in units of $\kappa$, removing distortions from the nonlinear current-frequency scaling and providing a physically meaningful representation where each unit corresponds to one resolvable TLS [\cref{fig: detection}(e)].
With the data standardized, we can apply a simple peak-counting method to identify candidate TLS.

\section{Detector calibration} \label{supp: calibration}

Detector calibration is initiated by determining the physical device parameters using a fit to a TLS-free `baseline' trace.
The baseline is chosen where residuals are flat, have no significant variation, and without any observable TLS couplings to ensure that the extracted physical parameters are as representative of the characteristic resonator response as possible.
The measurement noise level is then set by adding (white) noise such that the simulated residuals match the measured baseline residuals to within 1\%.
Using these fit parameters and the calibrated noise level, we generate large ensembles of synthetic resonator traces and compute their residuals.

\begin{figure}
    \centering
    \includegraphics{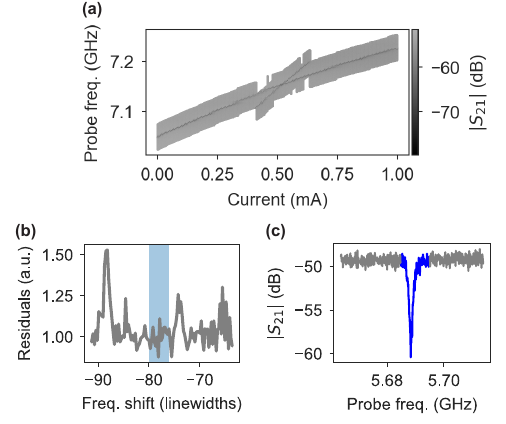}
    \caption{Data-filtering steps in the detection process. (a) Frequency collision of two nearby resonators. Occasionally resonator frequencies will overlap over the measured current span. Since the fitting algorithm cannot distinguish between the desired resonator and a transient overlap, such regions must be manually excluded from the analysis. (b) The blue highlighted region is suitable as the calibration data for that resonator. The calibration data must be clearly free of TLS peak signatures and be at the baseline of residuals. (c) The automated background filter. The algorithm separates the background (gray) from resonator response (blue).}
    \label{fig: filtering}
\end{figure}

To determine the standard characteristics of the resonator the experimenter selects a section of the dataset for calibration.
This calibration dataset is chosen based on the residuals: specifically, a region with no characteristic TLS peak and which lies at the constant residual baseline.
This indicates that deviations between the data and fit are primarily due to measurement noise, which is assumed to be Gaussian.
An example of this choice is shown in \cref{fig: filtering}(b).
The fit parameters of the calibration dataset are averaged and, with added noise, used to generate synthetic data closely resembling the resonator under investigation. Those same fit parameters also generate synthetic data that includes TLS coupling.
This forms the basis for the calibration of detection threshold in \cref{fig: detection}(d), and the computation of detector statistics.

Calibrating the TLS detections with synthetic data is crucial for ensuring an appropriate detection threshold and computing detector statistics and error.
As such, we need to ensure that the synthetic data is as faithful as possible to the real devices that it is simulating.
Our resonators are measured in the hanger configuration with response~\cite{khalil_analysis_2012,bruno_reducing_2015}

\begin{align}\label{eq: hanger}
    S_{21}(\omega)&=A\left(1+\alpha\frac{\omega-\omega_r}{\omega_r}\right)\\\nonumber
    &\times\left(1-\frac{\frac{Q_l}{|Q_e|}e^{i\theta}}{1+2iQ_l\frac{\omega-\omega_r}{\omega_r}}\right)e^{i\phi_v\omega+\phi_0}
\end{align}
for probe frequency $\omega/2\pi$, resonant frequency $\omega_r/2\pi$, off-resonant transmission amplitude $A$ and slope $\alpha$. 
The loaded quality factor is $Q_l$ where $1/Q_l=1/Q_c+1/Q_i$ for the coupling and internal quality factors respectively, $Q_e=|Q_e|e^{-i\theta}$ where $1/Q_c=Re(1/Q_e)$, and the phase slope and offset are $\phi_v$ and $\phi_0$ respectively, relating to the time-of-flight through the sample.

We modeled the hanger response in the presence of a TLS coupling as~\cite{petersson_circuit_2012,dejong_gatebased_2021}

\begin{align}\label{eq: hanger tls}
    S_{21}(\omega)&=1-\frac12\frac{\omega_r/Q_e}{i(\omega-\omega_r) + \frac{\omega_r}{2Q_l} + ig\chi(\omega)}\\
    \chi(\omega)&=g\frac{\langle\sigma_z\rangle}{\omega_{\text{TLS}}-\omega+i\langle\sigma_z\rangle\frac{\gamma_{\text{TLS}}}{2})}\nonumber
\end{align}
where $\omega_{\text{TLS}}$ is the TLS transition frequency and $\langle\sigma_z\rangle=\tanh\left(\frac{\hbar\omega_{\text{TLS}}}{k_BT}\right)$ is the average thermal population of the TLS. For compactness, the overall amplitude and phase factors appearing in \cref{eq: hanger} are omitted from \cref{eq: hanger tls} but are retained in the synthetic data generation.

We found that a good initial guess can significantly improve the quality of fits to the resonator response in \cref{eq: hanger}, so we employed a background filter to separate on-resonance and off-resonance $S_{21}$.
The background filter uses a Savitzky-Golay filter to smooth the data, estimates the derivative of the smoothed response, computes a moving-window variance, and then applies a second smoothing pass to identify the region of highest variation, which corresponds to the resonator response.
The filter allows us to fit the on-resonance and off-resonance components of the response separately, giving better estimates of the transmission slope, phase slope, and resonator linewidth.
The effect of the background filter is shown in \cref{fig: filtering}(c).

Once the calibration data has been fit, we add Gaussian noise to ensure the simulated residuals are realistic. 
We use the background filter once more to estimate the SNR of the data, taking the magnitude of the filtered resonator response and the standard deviation of the noise in the background. 
Because we are generating synthetic data from a model and adding noise artificially, the noisy synthetic data will be centered at the model. 
Though we can achieve high agreement between our fits and data, the fit will not be as perfectly centered in real data as the model is in synthetic data. 
This results in a slight (approximately 5\%) reduction in residuals in synthetic data. 
To remedy this and keep the synthetic residuals consistent with measured residuals, we iteratively increase the noise added to synthetic data, until there is agreement within 1\% between the synthetic and measured residuals.

\section{TLS density estimation and statistics}\label{supp: stats}
As with any threshold-based detector, it is necessary to consider the probabilities of both false positives and false negatives when assessing the reliability of our TLS identification.
During calibration with synthetic datasets, we define two base rates: the false negative probability ($fn$), representing the chance that a resonator trace with an actual mode splitting yields a residual below the detection threshold, and the false positive probability ($fp$), representing the chance that a trace without TLS coupling yields a residual above threshold.
These probabilities correspond to the area under the yellow curve to the left of the threshold and the violet curve to the right of the threshold, respectively, in \cref{fig: detection}(d).

However, these values do not directly represent the true false positive and false negative rates associated with our peak-based detection criteria.
Our detection method identifies TLS signatures as peaks within the residuals plotted against resonator frequency shift (measured in units of the linewidth $\kappa$).
To enable precise and uniform analysis, we interpolate the residuals such that the data are evenly spaced with a resolution of one-quarter linewidth, yielding five residual points per linewidth interval.
Before running the peak finder, we apply a Savitzky–Golay filter with window width 5 (corresponding to one linewidth) and polynomial order 1 to suppress noise-induced spurious peaks.

A detection is registered only when these five consecutive residual values satisfy the peak condition: the central value must lie above the detection threshold, and the two adjacent points on either side must decrease monotonically away from the center.
Thus, for a spurious peak to arise by chance in noise, not only must one or more values exceed the threshold, but they must also arrange themselves in a peak-like configuration.

\begin{figure}[t]
    \centering
    \includegraphics{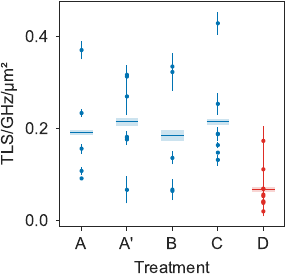}
    \caption{Expanded view of TLS densities across resonators. As in the main text, horizontal lines indicate the mean TLS density, and shaded rectangles denote the standard error of the mean. Dots represent individual resonator measurements, and vertical lines show the 68.27\% credible intervals.}
    \label{fig: tls densities combined}
\end{figure}

The probability that such a false peak forms purely by chance  defines the true false positive rate ($FP$), which is given by $\frac{6}{5!}(1-(1-fp)^5)=\frac{1-(1-fp)^5}{20}$.
Here $\frac{6}{5!}$ accounts for the six permutations (out of 120) in which five values can form a symmetric peak, and the remaining factor captures the likelihood that at least one of the five residuals exceeds the threshold.

The true false negative rate ($FN$) is $(fn)^5$: which reflects the probability that all five residual values lie below the detection threshold.
This expression assumes that every TLS coupling gives rise to a peak-like structure (i.e., a local maximum with neighboring values that decrease monotonically on either side) within the interpolated residuals.

To compute the true number of all detectable TLS, $N_T$, from the measured number of detections per resonator, $N_m$, we apply an empirical Bayesian framework. We assume that TLS occur independently across frequency bins of width equal to the resonator linewidth. 
Under this assumption, the number of true TLS $N_T$ follows a Poisson prior,
\begin{gather}
   P(N_T|\lambda)=\frac{e^{-\lambda}\lambda^{N_T}}{N_T!},
\end{gather}
where $\lambda$ is the mean TLS count in a given resonator.

In the measurement procedure, the observed detection number $N_m$ differs from $N_T$ due to false negatives and false positives. Let $N_n$ and $N_p$ denote the number of missed TLS and spurious detections, respectively.
These satisfy

\begin{gather}
N_m = N_T + N_p - N_n, \\
0 \leq N_n \leq N_T, \\
0 \leq N_p \leq B - N_T,
\end{gather}

where $B$ is the total number of resonator-linewidth-sized bins in the frequency sweep.
The above relations simply enforce the fact that at most one TLS can occur per bin.

Assuming independent detection outcomes, a true TLS is registered (missed) with probability $1-FN$ $(FN)$, while an empty bin produces (doesn't produce) a false positive (false negative) with probability $FP$ $(1-FP)$.
Combining these independent outcomes, the conditional probability of making $N_m$ detections given $N_T$ true TLS can be written as

\begin{multline}\label{eq: combinatorics}
    P(N_m | N_T) = \sum_{j=0}^{\min(N_m, N_T)} \Bigg[\binom{N_T}{j} (1 - FN)^j (FN)^{N_T - j} \\
    \times \binom{B - N_T}{N_m - j} (FP)^{N_m - j} (1 - FP)^{B - N_T - (N_m - j)}\Bigg].
\end{multline}

Next, the marginal likelihood for the Poisson rate parameter $\lambda$ is

\begin{align}
    P(N_m|\lambda) = \sum_{N_T=0}^B\text{Poisson}(N_T|\lambda)\times P(N_m|N_T).
\end{align}

We estimate $\lambda$ by maximizing this likelihood,
\begin{align}
    \lambda^*=\text{arg max}_\lambda\left[P(N_m|\lambda)\right].
\end{align}
Once this rate parameter is determined, the posterior distribution for the true number of TLS in the resonator is:

\begin{equation}
    P(N_T|N_m)\propto P(N_m|N_T)\times\text{Poisson}(N_T|\lambda^*).
\end{equation}
The inferred TLS density is then obtained from the posterior mean,
\begin{align}
    \rho=\frac{\sum_{N_T} N_T P(N_T|N_m)}{\Delta f\times A},
\end{align}

where $\Delta f$ is the frequency range of the resonator measurement and $A$ is the total junction area.
Finally, to present error bounds on $\rho$, we interpolate the discrete distribution $P(N_T|N_m)$ to form a normalized probability density function, from which we extract the 68.27\% credible interval for the true number of TLS coupled to each resonator.

In \cref{fig: tls densities}(a), the TLS densities for each resonator were shown without individual error bars, and only the error bars on the mean TLS density were displayed to maintain visual clarity.
In \cref{fig: tls densities combined} we show TLS densities for all resonators across devices, with error bars corresponding to 68.27\% credible intervals.
Specifically, for any device $\alpha$ containing $N_\alpha$ resonators, the mean TLS density $\rho_\alpha$ and the associated error bounds $\sigma_\alpha^{\pm}$ are given by,
\begin{align}
    \rho_\alpha&=\frac{1}{N_\alpha}\sum_i \rho_{\alpha,i}\\
    \sigma_\alpha^{\pm}&=\frac{1}{N_\alpha}\sqrt{\sum_i {\sigma^\pm_{\alpha,i}}^2}
\end{align}
where for each resonator \(i\) in device \(\alpha\), \(\sigma^+_{\alpha, i}\) and \(\sigma^-_{\alpha, i}\), represent the upper and lower limits of the credible interval for that resonator, respectively. 

\begin{table}[t!]
    \centering
    \begin{tabular}{c|c}
        Treatment & Shapiro-Wilk p-value \\
        \hline
        A + A\textsf{'} & 0.717 \\
        B & 0.092 \\
        C & 0.024 \\
        D & 0.028
    \end{tabular}
    \caption{Shapiro-Wilk p-values for each of the treatments under consideration. Treatments C and D exhibit statistically significant deviations from normality.}
    \label{tab: shapiro wilk}
\end{table}

\begin{table}[t!]
    \centering
    \begin{tabular}{c|c|c}
        Treatment 1 & Treatment 2 & Kruskal-Wallis p-value \\
        \hline
        A + A\textsf{'} & B & 0.833 \\
        A + A\textsf{'} & C & 0.866 \\
        A + A\textsf{'} & D & 0.001 \\
        B & C & 0.465 \\
        B & D & 0.028 \\
        C & D & 0.002 \\
    \end{tabular}
    \caption{Kruskal-Wallis p-values for all combinations of treatments. Treatments A and A\textsf{'} are grouped together because they are nominally the same, and we do not observe significant variations in the densities between those treatments. p-values between treatment D and all other treatments show statistically significant differences between the underlying distributions.}
    \label{tab: kruskal wallis}
\end{table}

\section{Comparison of different fabrication treatments} \label{supp: distribution tests}

To quantify the difference in TLS statistics of different devices we used a Shapiro-Wilk test~\cite{shapiro_analysis_1965a}. 
We found that devices C and D exhibited non-Gaussian behavior which may stem from limited detection counts or low false positive and false negative rates.
Given the non-normal behavior in some devices, we further applied the non-parametric Kruskal-Wallis test to compare distributions \cite{kruskal_use_1952}.
All statistical tests were evaluated at a significance level of $\alpha = 0.05$.
The results of these tests are summarized in \cref{tab: shapiro wilk} and \cref{tab: kruskal wallis}.

\section{Structural characterization}\label{supp: tem}

We fabricated cross-sectional TEM samples  using standard lift-out procedures using a Ga+ focused ion beam instrument (FEI Helios 600i Dual Beam FIB-SEM). Protective layers of amorphous carbon and Pt were used to minimize damage to the sample surface. A final milling voltage of 2 kV was used to reduce sample damage, and a cryo-can was used to minimize redeposition.
The samples were imaged in a Thermo Fisher Scientific Themis Z aberration-corrected STEM at 200 or 300 kV using a convergence angle of 25.2\,mrad.

To measure grain sizes, we made use of the fact that in BF-STEM images such as \cref{fig: stem}(a), electron channeling contrast leads to varying intensity between crystal grains based on their orientation to the electron beam. 
We use this contrast to measure the Al grain size in each device, obtaining 30-80 nm for devices A-C and 100-200 nm for device D.

In order to compare the relative impact of different morphological features, we use ridge regression to model the TLS density as a function of the metrics measured by STEM.
To mitigate the effects of multicolinearity on permutation importance, we utilized one representative feature from each cluster of correlated features~\cite{harrell_regression_2015}. 
Feature clusters were created by computing pairwise Spearman correlation coefficients, \(\rho\) between features and converting them into a distance metric, \(1 - |\rho|\). 
Hierarchical clustering was then performed using Ward linkage~\cite{wardjr._hierarchical_1963}. 
For each cluster, the representative feature was selected as the one most strongly correlated with TLS density using Spearman's correlation coefficient~\cite{dormann_collinearity_2013}. 
We define a threshold distance by calculating the Leave-One-Out-Cross-Validation (LOOCV) $R^2$ value for a ridge regression model using the clustering threshold, and choose the threshold which maximizes the LOOCV $R^2$ value for performing permutation importance.


%

\end{document}